\documentclass[12pt,prb,aps]{revtex4}
\usepackage{amsfonts}
\usepackage{amssymb}

\begin{document}
\title{Jacobi structures in $\mathbb{R}^3$}
\author{F. Haas\footnote{Electronic address: ferhaas@unisinos.br}}
\affiliation{UNISINOS - Ci\^encias Exatas e Tecnol\'ogicas \\ Av. Unisinos, 950\\
93022-000 - S\~ao Leopoldo, RS, Brazil}
\begin{abstract}
\noindent The most general Jacobi brackets in $\mathbb{R}^3$ are constructed after solving the equations imposed by the Jacobi identity. Two classes of Jacobi brackets were identified, according to the rank of the Jacobi structures. The associated Hamiltonian vector fields are also constructed. 
\end{abstract}
\maketitle
\section{Introduction}
Jacobi brackets have all properties of Poisson brackets, except from the fact that they are not necessarily derivations. A manifold endowed with a Jacobi bracket is called a Jacobi manifold. In this context, Jacobi manifolds are natural generalizations of Poisson, contact and locally conformal symplectic manifolds. Jacobi manifolds were introduced by Lichnerowicz \cite{Lichnerowicz} and, in a local Lie algebra setting, by Kirillov \cite{Kirillov} The general properties of Jacobi manifolds are discussed, for instance, in references 3 and 4 below. Some recent advances on the study of Jacobi manifolds can be found in references from 5 to 16.  

The present work is devoted to the explicit construction of Jacobi structures. Although contact and locally conformal symplectic manifolds are general concrete examples of Jacobi manifolds, there is a lack of knowledge of other possible classes of Jacobi brackets even for low dimensional manifolds. An exception in this regard is given by linear Jacobi structures on vector bundles \cite{Iglesias2}. This is to be compared with the Poisson manifolds case, where, in recent years, there has been much work for the explicit construction of Poisson structures, generalizing the well-known case of Lie-Poisson structures \cite{Kisisel}-\cite{Ibanez}. In particular, there have been intensive efforts in the derivation of new classes of three-dimensional Poisson structures, with application to three-dimensional dynamical systems \cite{Haas}-\cite{Plank2}. More recently, Ay {\it et al.} \cite{Ay} have found the general solution of the determining equation for Poisson structures in $\mathbb{R}^3$. The purpose of the present work is to extend this result, obtaining the general form of Jacobi structures in $\mathbb{R}^3$. Our approach is not applicable to generic three-dimensional manifolds. 

The paper is organized as follows. In Section II, we briefly review the basic definitions about Jacobi manifolds. In Section III, we consider the specific case of Jacobi structures in $\mathbb{R}^3$. The determining equations for Jacobi brackets in $\mathbb{R}^3$ are obtained and solved. In Section IV, the associated Hamiltonian vector fields are discussed. Section V is reserved to our conclusions.

\section{Jacobi manifolds}
Here we review the essential results about Jacobi manifolds. More detailed accounts on the subject can be found in references from 1 to 4. By definition, a Jacobi structure on a manifold $M$ is a 2-vector ${\bf\Lambda}$ and a vector field ${\bf E}$ on $M$ such that 
\begin{equation}
\label{e0}
[{\bf\Lambda}, {\bf\Lambda}] = 2 {\bf E} \wedge {\bf\Lambda} \,, \quad [{\bf E}, {\bf\Lambda}] = 0 \,,
\end{equation}
where $[\,,]$ is the Schouten-Nijenhuis bracket \cite{Olver}. Let $C^{\infty}(M,\mathbb{R})$ be the algebra of $C^\infty$ real-valued functions on $M$. If $(M, {\bf\Lambda}, {\bf E})$ is a Jacobi structure, 
then the space $C^{\infty}(M,\mathbb{R})$ endowed with a mapping $\{\,,\}: C^{\infty}(M,\mathbb{R}) \times C^{\infty}(M,\mathbb{R}) \rightarrow C^{\infty}(M,\mathbb{R})$ becomes a local Lie algebra in the sense of Kirillov \cite{Kirillov}. This so-called Jacobi bracket $\{\,,\}$ is defined by
\begin{equation}
\label{jb}
\{f,g\} = {\bf\Lambda}(df,dg) + f {\bf E}(g) - g {\bf E}(f) \,,
\end{equation}
for all $f, g \in C^{\infty}(M,\mathbb{R})$. The Jacobi bracket is $\mathbb{R}$-bilinear, skew-symmetric and satisfies the Jacobi identity. In other words, 
\begin{eqnarray}
\{c_{1} f + c_{2} g, h\} &=& c_{1} \{f,h\} + c_{2} \{g,h\} \,, \nonumber \\
\{f,g\} &=& - \{g,f\} \,, \\
\{\{f,g\},h\} &+& \{\{g,h\},f\} + \{\{h,f\},g\} = 0 \,, \nonumber
\end{eqnarray}
for all $f, g, h \in C^{\infty}(M,\mathbb{R})$ and $c_{1}, c_{2} \in \mathbb{R}$. The local Lie algebra character of $C^{\infty}(M,\mathbb{R})$ is assured by 
\begin{equation}
\rm{support} \, \{f,g\} \quad \subseteq \quad \rm{support} \, f \quad \cap \quad \rm{support} \, g \,, 
\end{equation}
for all $f, g \in C^{\infty}(M,\mathbb{R})$. 

In addition, the Jacobi bracket is a first-order differential operator in each of its arguments using ordinary multiplication of functions, 
\begin{equation}
\label{e1}
\{f g,h\} = f\{g,h\} + \{f,h\}g - f g \,\{1,h\} \,,
\end{equation}
for all $f, g, h \in C^{\infty}(M,\mathbb{R})$. As apparent from (\ref{e1}), the Jacobi bracket is not a derivation unless the constant unit function has a vanishing Jacobi bracket with all functions in $C^{\infty}(M,\mathbb{R})$. It happens if and only if the vector field ${\bf E}$ vanish, ${\bf E} \equiv 0$. If ${\bf E} \equiv {\bf 0}$, then $(M, {\bf\Lambda})$ is a Poisson manifold. In this sense, Jacobi manifolds are generalizations of Poisson manifolds. Examples of manifolds with Jacobi but not Poisson structures are contact and locally conformal symplectic manifolds. 

As shown by Lichnerowicz \cite{Lichnerowicz}, any Jacobi structure $(M, {\bf\Lambda}, {\bf E})$ can be associated to a higher-dimensional Poisson structure $({\bf\Pi}, M \times \mathbb{R})$, defined by 
\begin{equation}
\label{e2}
{\bf\Pi}({\bf x},t) = \exp(-t)\left({\bf\Lambda}({\bf x}) + \frac{\partial}{\partial t}\wedge {\bf E}\right) \,,
\end{equation}
where $(x,t)$ are local coordinates in $({\bf\Pi}, M \times \mathbb{R})$. While equation (\ref{e2}) provides a simple recipe to obtain a Poisson structure from a Jacobi structure in a lower dimensional manifold, it is not trivial to construct a Jacobi structure in the manifold $M$ itself. Hence the importance of deriving concrete examples of Jacobi structures in manifolds like $\mathbb{R}^3$, for instance.

In a local coordinate chart $x^{i}$, for $i = 1,\dots,n$ with $n = \rm{dim} M$, we have the following expressions for the tensor field ${\bf\Lambda}$, the vector field ${\bf E}$ and the Jacobi bracket of two functions $f$ and $g$, 
\begin{eqnarray}
{\bf\Lambda} &=& \frac{1}{2}\Lambda^{ij}\,\partial_{i}\wedge\partial_{j} \,, \quad {\bf E} = E^{i}\partial_{i} \,,  \\
\{f,g\} &=& \Lambda^{ij}\,\partial_{i} f \, \partial_{j} g + f \,E^{i}\,\partial_{i} g - g \,E^{i}\,\partial_{i} f \,. \nonumber
\end{eqnarray}
The summation convention was used, as well as the notation $\partial_{i} = \partial/\partial x^{i}$. In addition, the Schouten-Nijenhuis bracket conditions (\ref{e0}) traduces into
\begin{eqnarray}
\label{j1}
\Lambda^{im}\partial_{m}\Lambda^{jk} &+& \Lambda^{jm}\partial_{m}\Lambda^{ki} + \Lambda^{km}\partial_{m}\Lambda^{ij} + \Lambda^{ij}E^k + \Lambda^{jk}E^i + \Lambda^{ki}E^{j} = 0 \,, \\ \label{j2}
E^{k}\partial_{k}\Lambda^{ij} &-& \Lambda^{ik}\partial_{k}E^j + \Lambda^{jk}\partial_{k}E^{i} = 0 \,.
\end{eqnarray}
As can be verified, equations (\ref{j1}-\ref{j2}) are completely equivalent to the Jacobi identity for the Jacobi bracket (\ref{jb}). In this context, we call (\ref{j1}-\ref{j2}) the Jacobi equations. The general solution for the Jacobi equations yields the general form of the Jacobi structures in $\mathbb{R}^n$. Ay {\it et al.} have solved the Jacobi equations for $n = 3$ and ${\bf E} = {\bf 0}$, that is, the Poisson case in $\mathbb{R}^3$. In the next Section we allow for non-vanishing vector fields ${\bf E}$, looking for the general class of Jacobi structures in $\mathbb{R}^3$.

\section{Jacobi structures in $\mathbb{R}^3$} 
Consider the $\mathbb{R}^3$ case, introducing a vector field ${\bf A} = A^{i}\partial_{i}$ according to 
\begin{equation}
\label{e5}
\Lambda^{ij} = \varepsilon_{ijk}A^{k} \,,
\end{equation}
using the Levi-Civita symbol $\varepsilon_{ijk}$. In other words, ${\bf A} = (\Lambda^{23},\Lambda^{31},\Lambda^{12})$. In terms of this vector field ${\bf A}$, the Jacobi equations (\ref{j1}-\ref{j2}) are rewritten as 
\begin{eqnarray}
\label{jj1}
{\bf A}\cdot(\nabla\times{{\bf A}} - {\bf E}) &=& 0 \,,\\
\label{jj2}
{\bf E}\times(\nabla\times{{\bf A}}) + {\bf A}\,\nabla\cdot{{\bf E}} &=& \nabla({\bf A}\cdot{{\bf E}}) \,,
\end{eqnarray}
using standard symbols of vector calculus in $\mathbb{R}^3$. 

Using (\ref{jj1}-\ref{jj2}) we can proceed to construct Jacobi structures using the well-known language of vector calculus in $\mathbb{R}^3$. Before we do that, it is interesting to interpret geometrically equation (\ref{jj1}). Consider $(M,{\bf\Lambda},{\bf E})$ a given Jacobi structure and let ${\bf\Lambda}^\sharp : T^{*}M \rightarrow TM$ be the vector bundle map associated with ${\bf\Lambda}$. In other words, for all ${\bf p} \in M$, $\zeta, \eta \in T_{p}^{*}M$, 
\begin{equation}
{\bf\Lambda}(\zeta,\eta) = \left\langle \eta, {\bf\Lambda}^{\sharp}(\zeta)\right\rangle \,,
\end{equation}
where $\langle\,, \rangle$ denotes the natural pairing. In terms of the vector field ${\bf A}$, the last equation shows that the image of the vector bundle map $\Lambda^\sharp$ is the plane orthogonal to ${\bf A}$. Taking into account (\ref{jj1}), written as ${\bf A}\cdot{\bf E} = {\bf A}\cdot\nabla\times{\bf A}$, we conclude that the condition ${\bf A}\cdot\nabla\times{\bf A} \neq 0$ would imply that the vector ${\bf E}$ does not belong to the image of ${\bf\Lambda}^\sharp$, or, in other words, that the rank of the Jacobi structure is everywhere 3. On the other hand, the condition ${\bf A}\cdot\nabla\times{\bf A} \equiv 0$ would imply that the vector ${\bf E}$ takes its values in the image of ${\bf\Lambda}^\sharp$. Together with ${\bf A} \neq 0$, that means that the rank of the Jacobi structure would be everywhere 2. 

Now consider the calculation of Jacobi structures. For ${\bf A} \equiv 0$, equations (\ref{jj1}-\ref{jj2}) are identically satisfied for arbitrary ${\bf E}$, but this will not produce very interesting Jacobi structures. Disregarding the too trivial case ${\bf A} \equiv 0$, we conclude that equation (\ref{jj1}) is equivalent to 
\begin{equation}
\label{e8}
{\bf E} = \nabla\times{{\bf A}} - {\bf g}\times{{\bf A}} \,, 
\end{equation}
for some vector field ${\bf g}$ in $\mathbb{R}^3$, to be determined. The Poisson case ${\bf E} = {\bf 0}$ is obtained for 
\begin{equation}
\label{p}
\nabla\times{{\bf A}} = {\bf g}\times{{\bf A}} \,.
\end{equation}
As shown by Ay and coworkers in a different notation \cite{Ay}, the general solution for (\ref{p}) is given by 
\begin{equation}
\label{ay}
{\bf A} = \mu\nabla\psi \,,
\end{equation}
for $\mu, \psi \in C^{\infty}(\mathbb{R}^{3},\mathbb{R})$. The solution (\ref{ay}) has shown to be valid \cite{Ay} also in the neighborhood of some classes of irregular points, where ${\bf A} = {\bf 0}$. The corresponding vector field ${\bf g}$ is given by
\begin{equation}
{\bf g} = \frac{\nabla\mu}{\mu} \,.
\end{equation}

Returning to the not necessarily Poisson case and substituting the form (\ref{e8}) into (\ref{jj2}), we get the following condition, 
\begin{equation}
\label{c}
\nabla({\bf A}\cdot\nabla\times{{\bf A}}) = {\bf g}\,({{\bf A}}\cdot\nabla\times{{\bf A}}) - {\bf A}\,({\bf A}\cdot\nabla\times{\bf g}) \,.
\end{equation}

In the remaining part of the Section we examine the general solution for (\ref{c}), thus determining the possible classes of Jacobi structures in $\mathbb{R}^3$. We consider two cases, according to the rank of the Jacobi structure.

\subsection{Rank $3$ Jacobi structures}
Suppose 
\begin{equation}
\label{r3}
{\bf A}\cdot\nabla\times{{\bf A}} \neq 0 \,.
\end{equation}
In this situation, $(\mathbb{R}^{3},{\bf\Lambda})$ is certainly not a Poisson structure, with $2$-vector ${\bf\Lambda}$ specified by (\ref{e5}). However, we can construct Jacobi structures defining a vector field ${\bf h}$ such that 
\begin{equation}
\label{e10}
{\bf g} = \nabla\phi + {\bf h} \,,
\end{equation}
for 
\begin{equation}
\label{e11}
\phi = \ln\left({\bf A}\cdot\nabla\times{{\bf A}}\right) \,.
\end{equation}
Of course the function $\phi$ is not well defined unless ${\bf A}\cdot\nabla\times{{\bf A}}$ is not identically vanishing. Inserting (\ref{e10}) into (\ref{c}), the result is 
\begin{equation}
\label{e12}
{\bf h}\,({\bf A}\cdot\nabla\times{{\bf A}}) - {\bf A}\,({\bf A}\cdot\nabla\times{\bf h}) = 0 \,.
\end{equation}
Cross product with of (\ref{e12}) with ${\bf A}$ gives 
\begin{equation}
\label{e13}
{\bf A}\times{\bf h} = {\bf 0} \rightarrow {\bf h} = \alpha\,{{\bf A}} \,,
\end{equation}
for an arbitrary real-valued function $\alpha$ on $\mathbb{R}^{3}$. We can take $\alpha \equiv 0$ without loss of generality, since, according to (\ref{e8}) and (\ref{e10}), this function will add nothing to the vector field ${\bf E}$. Hence, in the rank 3 case the determining equation (\ref{c}) is solved for any vector field ${\bf A}$ satisfying ${\bf A}\cdot\nabla\times{{\bf A}} \neq 0$, taking ${\bf g} = \nabla\phi$ with $\phi$ given by (\ref{e11}). The corresponding vector field ${\bf E}$ follows from (\ref{e8}). After using some vector identities, the result is 
\begin{equation}
\label{e14}
{\bf E} = e^{\phi}\,\,\nabla\times\left(e^{-\phi}\,{\bf A}\right) \,.
\end{equation}

Using (\ref{jb}), (\ref{e5}) and the above solution ${\bf E}$, we get the Jacobi bracket between any functions $f, g \in C^{\infty}(\mathbb{R}^{3}, \mathbb{R})$, 
\begin{equation}
\label{e15}
\{f,g\} = {\bf A}\cdot\nabla\,f\times\nabla\,g + e^{\phi}\,(f\nabla\,g - g\nabla\,f)\cdot\nabla\times(e^{-\phi}\,{\bf A}) \,.
\end{equation}
It takes a simpler form for $\phi = \rm{constant}$, in which case
\begin{equation}
\{f,g\} = {\bf A}\cdot\nabla\,f\times\nabla\,g + (f\nabla\,g - g\nabla\,f)\cdot\nabla\times\,{\bf A} \,.
\end{equation}
In conclusion, the Jacobi bracket (\ref{e15}) has three free ingredients, namely the three components of the vector field ${\bf A}$, as long as ${\bf A}\cdot\nabla\times{{\bf A}}$ is not identically null. For simplicity, we do not consider the behavior of the solution in the neighborhood of points into domains $\Omega \subset \mathbb{R}^3$ where ${\bf A}\cdot\nabla\times{{\bf A}} = 0$.

On $\mathbb{R}^3$, rank 3 Jacobi structures are in one to one correspondence with contact 1-forms. Specifically, giving a contact 1-form $\theta$ on $\mathbb{R}^3$ such that $\theta \wedge d\theta \neq 0$, there is a Jacobi structure $(\mathbb{R}^{3}, {\bf\Lambda}, {\bf E})$ satisfying ${\bf\Lambda} \wedge {\bf E} \neq 0$ and $i_{\theta}({\bf\Lambda}) = 0$, $i_{\bf E}(\theta) = 1$. In the case of the rank 3 Jacobi structures of this Section, using cartesian coordinates  we can show that 
\begin{equation}
{\bf\Lambda} \wedge {\bf E} = {\bf A}\cdot{\bf E} \, \frac{\partial}{\partial x} \wedge \frac{\partial}{\partial y} \wedge \frac{\partial}{\partial z} = {\bf A}\cdot\nabla\times{\bf A} \, \frac{\partial}{\partial x} \wedge \frac{\partial}{\partial y} \wedge \frac{\partial}{\partial z} \neq 0 \,. 
\end{equation}
Since ${\bf\Lambda} \wedge {\bf E}$ is nowhere zero, there is a contact 1-form $\theta$ associated to this Jacobi structure. It can be shown that 
\begin{equation}
\label{1f}
\theta = e^{-\phi} A^{i} dx^i \,.
\end{equation}
satisfies 
\begin{eqnarray}
i_{\theta}({\bf\Lambda}) &=& - e^{-\phi} ({\bf A}\times{\bf A})\cdot\nabla = 0 \,, \\
i_{\bf E}(\theta) &=& e^{-\phi}{\bf A}\cdot(\nabla\times{\bf A} - \nabla\phi\times{\bf A}) = e^{-\phi} {\bf A}\cdot\nabla\times{\bf A} = 1 \,,
\end{eqnarray}
the last equality following from the definition of $\phi$. Moreover,  
\begin{equation}
\theta \wedge d\theta = e^{-\phi} \, dx \wedge dy \wedge dz \,,
\end{equation}
which never vanishes. Therefore, the 1-form $\theta$ given by (\ref{1f}) qualifies as the contact 1-form associated to rank 3 Jacobi structures in $\mathbb{R}^3$. As a corollary, we conclude that 
\begin{equation}
d\theta = e^{-\phi}\,(\partial_{i} A^j - A^{j}\partial_{i}\phi) \, dx^{i} \wedge dx^j 
\end{equation}
is a symplectic structure on the vector bundle $\rm{ker}\,\theta \rightarrow \mathbb{R}^3$ defined as the set of vector fields orthogonal to ${\bf A}$ at every point.

Another way to interpret the above construction is in terms of the characteristic distribution \cite{Kirillov}-\cite{Dazord} of the Jacobi structure $(\mathbb{R}^{3}, {\bf\Lambda}, {\bf E})$, defined as the subbundle $D$ of $T(\mathbb{R}^{3})$ spanned by the set of all Hamiltonian vector fields. In other words, $D_{\bf p} = \rm{Span}\{{\bf\Lambda}^{\sharp}(\zeta)({\bf p}), {\bf E}({\bf p}), \forall \zeta \in T^{*}(\mathbb{R}^{3})\}$ is the fiber at a point ${\bf p} \in \mathbb{R}^{3}$. The characteristic distribution of a Jacobi structure is completely integrable, defining a foliation whose leaves are contact manifolds or locally conformal symplectic manifolds. The leaves of the foliation can be represented as the level sets of a function $f(x,y,z)$. In terms of the vector fields ${\bf A}$ and ${\bf E}$, the determining equations of the foliation are given by 
\begin{eqnarray}
\label{l1}
{\bf A}\times\nabla\,f &=& 0 \,, \\
\label{l2}
{\bf E}\cdot\nabla\,f &=& 0 \,.
\end{eqnarray}
For the rank 3 Jacobi structures of this Section, inserting (\ref{e14}) into (\ref{l2}) and considering equation (\ref{l1}) gives 
\begin{equation}
\nabla\times{\bf A}\cdot\nabla\,f = 0 \,,
\end{equation}
which is redundant since $\nabla\times{\bf A}\cdot\nabla\,f = \nabla\cdot({\bf A}\times\nabla\,f) = 0$ in virtue of (\ref{l1}). Hence, for rank 3 Jacobi structures in $\mathbb{R}^{3}$ the vector field ${\bf A}$ is always normal to the leaves of the foliation, as stated in (\ref{l1}). 

\subsection{Rank $2$ Jacobi structures}
For
\begin{equation}
\label{epp}
{\bf A}\cdot\nabla\times{{\bf A}} \equiv 0 \,,
\end{equation}
the determining equation (\ref{c}) simplifies to
\begin{equation}
\label{se}
{\bf A}\cdot\nabla\times{\bf g} = 0 \,,
\end{equation}
excluding the trivial case ${\bf A} = {\bf E} = {\bf 0}$. As shown by Ay and coworkers \cite{Ay}, the general solution for (\ref{epp}) is given in terms of two scalar functions $\mu$ and $\psi$, as in equation (\ref{ay}). Using (\ref{ay}), the equation (\ref{se}) traduces into 
\begin{equation}
\nabla\psi\cdot\nabla\times{\bf g} = \nabla\cdot({\bf g}\times\nabla\psi) = 0 \,,
\end{equation}
showing that the vector field ${\bf g}\times\nabla\psi$ is solenoidal.  Hence, there are real smooth functions $\xi_1$ and $\xi_2$ on $\mathbb{R}^3$ such that 
\begin{equation}
\label{x}
{\bf g}\times\nabla\psi = \nabla\xi_{1}\times\nabla\xi_{2} 
\end{equation}
in a neighborhood $U$ of every regular point of ${\bf g}\times\nabla\psi$. We exclude the trivial case $\nabla\psi = 0$. 

Inversion of (\ref{x}) gives
\begin{equation}
\label{gg}
{\bf g} = \frac{\nabla\psi\times(\nabla\xi_{1}\times\nabla\xi_{2})}{|\nabla\psi|^2} + \alpha\nabla\psi \,,
\end{equation}
where $\alpha$ is an arbitrary real-valued smooth function on $\mathbb{R}^{3}$ and 
\begin{equation}
\label{xx}
\nabla\psi\cdot\nabla\xi_{1}\times\nabla\xi_2 = 0 
\end{equation}
for consistency with (\ref{x}). Equation (\ref{xx}) imply a functional dependence between $\psi$, $\xi_1$ and $\xi_2$,  
\begin{equation}
\psi = \psi(\xi_{1},\xi_{2}) \,.
\end{equation}

Since we have solved the determining equation (\ref{c}), we can construct the vector field ${\bf E}$, using equations (\ref{e8}), (\ref{ay}) and (\ref{gg}). This procedure gives 
\begin{equation}
\label{ui}
{\bf E} = \nabla\mu\times\nabla\psi - \mu\nabla\xi_{1}\times\nabla\xi_2 \,.
\end{equation}
The corresponding Jacobi bracket reads
\begin{equation}
\label{jab}
\{f,g\} = \mu\nabla\psi\cdot\nabla{f}\times\nabla{g} + (f\nabla{g} -  g\nabla{f})\cdot(\nabla\mu\times\nabla\psi - \mu\nabla\xi_{1}\times\nabla\xi_{2})
\end{equation}
Notice that the real function $\alpha$ in (\ref{gg}) does not appear in the final form of the Jacobi bracket. Therefore, the Jacobi structure described by (\ref{jab}) has only four ingredients, namely $\mu$, $\xi_{1}$, $\xi_{2}$ and $\psi(\xi_{1},\xi_{2})$. 

The presence of the function $\mu$ in the Jacobi bracket can be eliminated by a conformal transformation. Indeed \cite{Vaisman2}, if $(M, {\bf\Lambda}, {\bf E})$ is a Jacobi structure in a manifold $M$, then $(M, \tilde{\bf\Lambda}, \tilde{\bf E})$ is also a Jacobi structure in $M$, where $\tilde{\bf\Lambda}$ and $\tilde{\bf E}$ are obtained from the conformal change
\begin{equation}
\tilde{\bf\Lambda} = \lambda {\bf\Lambda} \,, \quad \tilde{\bf E} = \lambda{\bf E} + {\bf\Lambda}^{\sharp}(d\lambda) \,, 
\end{equation}
for any $\lambda \in C^{\infty}(\mathbb{R}^{3},\mathbb{R})$. Applying such a conformal transformation to everywhere rank 2 Jacobi structures in $\mathbb{R}^3$ using $\lambda = 1/\mu$ gives 
\begin{equation}
\tilde{\Lambda}^{ij} = \varepsilon_{ijk}\,\partial_{k}\psi \,, \quad \tilde{E} = - \nabla\xi_{1}\times\nabla\xi_2 \,,
\end{equation}
with no presence of the arbitrary function $\mu$. However, this does not means that $\mu$ is to be taken as irrelevant in applications for dynamical systems, for instance.

A Jacobi structure everywhere of rank 2 on $\mathbb{R}^3$ determines a foliation of  $\mathbb{R}^3$ whose leaves are locally conformal symplectic surfaces \cite{Kirillov}-\cite{Dazord}. Perhaps the simplest way to describe such a foliation is in terms of the characteristic distribution of the Jacobi structure, as following from (\ref{l1}-\ref{l2}). Representing, as before, the leaves of the foliation as the level sets of a function $f(x,y,z)$ and using (\ref{ay}) and (\ref{ui}) in (\ref{l2}), we get 
\begin{equation}
{\bf E}\cdot\nabla\,f = - \mu \nabla\xi_{1}\times\nabla\xi_{2}\cdot\nabla\,f \equiv 0 \,,
\end{equation}
showing that $f = f(\xi_{1},\xi_{2})$. When we insert this result into (\ref{l1}) with ${\bf A}$ given by (\ref{ay}), we conclude that 
\begin{equation}
\frac{\partial\psi}{\partial\xi_1}\frac{\partial\,f}{\partial\xi_2} - \frac{\partial\psi}{\partial\xi_2}\frac{\partial\,f}{\partial\xi_1} = 0 \,,
\end{equation}
showing that $f$ and $\psi$ are functionally dependent. In conclusion, for rank 2 Jacobi structures in $\mathbb{R}^3$ there exists a locally conformal symplectic foliation whose leaves can be represented by the level sets of the function $\psi(x,y,z)$ entering the Jacobi bracket. 

The degenerate character of rank 2 Jacobi structures in $\mathbb{R}^3$ allows to search for Casimir functions associated to the Jacobi bracket. These Casimir functions can be defined in analogy with the Casimir functions of degenerate Poisson manifolds. In other words, we define Casimir functions as the non constant functions having a vanishing Jacobi bracket with any other function on $\mathbb{R}^3$. Let a Casimir function be denoted by $C = C(x,y,z)$. In terms of the vector fields ${\bf A}$ and ${\bf E}$, we have 
\begin{equation}
\label{cas}
\{f,C\} = (\nabla\,C\times{\bf A} - C {\bf E})\cdot\nabla{f} + f {\bf E}(C) \equiv 0 
\end{equation}
for an arbitrary smooth function $f \in C^{\infty}(\mathbb{R}^{3},\mathbb{R})$. Since $f$ is arbitrary, equation (\ref{cas}) is decomposed in two parts, 
\begin{eqnarray}
\label{cas1}
\nabla\,C\times{\bf A} = C{\bf E} \,, \\
\label{cas2}
{\bf E}(C) = 0 \,.
\end{eqnarray}
Observe that the Casimirs are preserved by the flow of the vector field ${\bf E}$. In addition, notice that, for $C\neq 0$, equation (\ref{cas2}) follows from (\ref{cas1}) taking the scalar product with $\nabla\,C$. Therefore, (\ref{cas1}) is sufficient for the construction of the Casimirs. T

For $C \neq 0$, scalar product of (\ref{cas1}) with ${\bf A}$ implies 
\begin{equation}
{\bf A}\cdot{\bf E} = 0 \,,
\end{equation}
incidentally the condition for rank 2 Jacobi structures in $\mathbb{R}^3$. Therefore, non trivial Casimirs can exist only in the case of rank 2 Jacobi structures, as expected. As will be explicitly shown in what follows, the condition (\ref{cas1}) is not only necessary but also sufficient for the existence of non trivial Casimirs. 

Equation (\ref{cas1}), written in terms of ${\bf A}$ as given in (\ref{ay}) and ${\bf E}$ as given in (\ref{ui}), reads, after some simple algebra considering $\mu \neq 0$, 
\begin{equation}
\label{ga}
\nabla\Gamma \times \nabla\psi = - \nabla\xi_{1}\times\nabla\xi_2 \,,
\end{equation}
where
\begin{equation}
\label{gam}
\Gamma = \ln\left(\frac{C}{\mu}\right) \,.
\end{equation}
After solving (\ref{ga}) for $\Gamma$, the Casimirs follows trivially from (\ref{gam}).

Scalar product of (\ref{ga}) with $\nabla\Gamma$ gives 
\begin{equation}
\nabla\Gamma\cdot\nabla\xi_{1}\times\nabla\xi_2 = 0 \,,
\end{equation}
showing that 
\begin{equation}
\Gamma = \Gamma(\xi_{1},\xi_{2}) \,.
\end{equation}
Inserting this information on the functional dependence of $\Gamma$ in (\ref{cas1}), we get
\begin{equation}
\left(\frac{\partial\Gamma}{\partial\xi_1} \frac{\partial\psi}{\partial\xi_2} - \frac{\partial\Gamma}{\partial\xi_2} \frac{\partial\psi}{\partial\xi_1} + 1\right) \nabla\xi_{1}\times\nabla\xi_2 = 0 \,.
\end{equation}
Since this equation holds for arbitrary $\xi_{1}$, $\xi_2$, we conclude that 
\begin{equation}
\label{carac}
\frac{\partial\Gamma}{\partial\xi_1} \frac{\partial\psi}{\partial\xi_2} - \frac{\partial\Gamma}{\partial\xi_2} \frac{\partial\psi}{\partial\xi_1} = - 1 \,.
\end{equation}
Equation (\ref{carac}) can be solved by the method of characteristics. The characteristic equations can be written as
\begin{equation}
\frac{d\xi_1}{\partial\psi/\partial\xi_2} = - \frac{d\xi_2}{\partial\psi/\partial\xi_1} = - d\Gamma \,.
\end{equation}
One of the characteristics is readily identified as $\psi(\xi_{1},\xi_{2})$. Without loss of generality, suppose $\partial\psi/\partial\xi_1 \neq 0$ in some neighborhood. Then, the inverse function theorem allows to consider $\xi_1 = \xi_{1}(\xi_{2};\psi)$, written locally as a function of $\xi_2$ and $\psi$. This allows writing the general solution for (\ref{carac}) according to
\begin{equation}
\label{crc}
\Gamma = \int_{\xi_{1}=\xi_{1}(\xi_{2};\psi)}\,\frac{d\xi_2}{\partial\psi/\partial\xi_1} + \bar{\Gamma}(\psi) \,,
\end{equation}
where $\bar\Gamma$ is an arbitrary function of the indicated argument. 

Equation (\ref{crc}), together with (\ref{gam}), provides a recipe to compute the Casimirs of a given rank 2 Jacobi structure in $\mathbb{R}^3$. Perhaps it would be interesting to provide a concrete example of the procedure. Consider a rank 2 Jacobi structure in $\mathbb{R}^3$ defined by 
\begin{equation}
\xi_1 = x \,, \quad \xi_2 = y \,, \quad \psi = x^2 + y^2 \,, \quad \mu = 1 \,.
\end{equation}
With (\ref{crc}) and then (\ref{gam}), we obtain the Casimirs in the form
\begin{equation}
C = \bar{C}(r^2)\,e^{\theta/2} \,,
\end{equation}
using cylindrical coordinates and defining 
\begin{equation}
\bar{C}(\psi) = e^{\bar{\Gamma}(\psi)} \,.
\end{equation}
In addition, we observe that in this case the leaves of the foliation of $\mathbb{R}^3$ in locally conformal symplectic surfaces are given by the cylinders defined by $\psi = x^2 + y^2 = {\rm constant}$. 

In the next Section, we discuss the Hamiltonian vector fields associated to the two types of Jacobi structures derived. 

\section{Hamiltonian vector fields associated to Jacobi structures in $\mathbb{R}^3$}
For a function $H \in C^{\infty}(M,\mathbb{R})$, the Hamiltonian vector field ${\bf v}_H$ associated with $H$ is defined by \cite{Lichnerowicz}
\begin{equation}
\label{h}
{\bf v}_H = {\bf\Lambda}^{\sharp}(dH) + H {\bf E} \,.
\end{equation}
In this context, the function $H$ is said to be the Hamiltonian for the vector field ${\bf v}_H$. In particular, 
\begin{equation}
{\bf v}_1 = {\bf E} \,,
\end{equation}
that is, the unit function $H = 1$ is a Hamiltonian associated to any vector field ${\bf E}$. In this sense, any dynamical system in a manifold $M$ can be viewed as a Hamiltonian system with Jacobi structure given by Hamiltonian function $H = 1$, vector field ${\bf E}$ being the dynamical vector field itself and tensor field ${\bf\Lambda} \equiv 0$. In addition, 
\begin{equation}
{\bf v}_{\{f,g\}} = [{\bf v}_{f},{\bf v}_{g}] \,,
\end{equation}
for any $f, g \in C^{\infty}(M,\mathbb{R})$, so that the mapping which associates a function with the corresponding Hamiltonian vector field is a Lie algebra homomorphism. Finally, notice that the Hamiltonian is not always a constant of motion, even if time-independent. In fact, for a time-independent Hamiltonian function, 
\begin{equation}
\frac{dH}{dt} = H {\bf E}(H) \,.
\end{equation}
Hence, $H$ is a constant of motion if and only if it is preserved by the flow associated to the vector field ${\bf E}$. 

In a local coordinate chart $x^{i}$, for $i = 1,\dots,n$ with $n = \rm{dim} M$, the Hamiltonian vector field at (\ref{h})
and the corresponding Hamilton equations in $M$ are then
\begin{equation}
\dot{x}^i = v_{H}^i = \Lambda^{ij}\partial_{j} H + H E^{i} \,,
\end{equation}
where ${\bf v}_H = v_{H}^{i}\partial_i$.  For $\mathbb{R}^3$ and in terms of the vector field ${\bf A}$, the result is 
\begin{equation}
\label{lc}
\dot{\bf r} = \nabla{H}\times{\bf A} + H{\bf E} \,,
\end{equation}
where ${\bf r} = (x,y,z)$. The expression (\ref{lc}) can be used to compute the Hamiltonian vector fields associated to the Jacobi structures of Section III. The two classes of Jacobi structures are treated separately. 

\subsection{Hamiltonian vector fields associated to rank $3$ Jacobi structures}
The class of Jacobi structures described in subsection III.A yields the following Hamilton equations, taking ${\bf E}$ as in (\ref{e14}), 
\begin{equation}
\label{s}
\dot{\bf r} = \nabla\times(H {\bf A}) + H {\bf A}\times\nabla\phi \,,
\end{equation}
with Hamiltonian $H$ and $\phi$ as in (\ref{e11}).  

At first sight, it appears that a simple rescaling ${\bf A} \rightarrow \bar{{\bf A}} = H {\bf A}$ would be sufficient to incorporate $H$ into the new vector field $\bar{{\bf A}}$, eliminating one irrelevant function. However, the dependence of $\phi$ on ${\bf A}$ prevents this possibility. Indeed, consider $\bar{\bf A}$ and $\bar{\phi}$ given by 
\begin{equation}
\bar{{\bf A}} = H {\bf A} \,, \quad \bar\phi = \ln(\bar{{\bf A}}\cdot\nabla\times\bar{{\bf A}}) \,.
\end{equation}
With these definitions, (\ref{s}) reads
\begin{equation}
\label{ss}
\dot{\bf r} = \nabla\times\bar{{\bf A}} + \bar{{\bf A}}\times\nabla\bar{\phi} + 2 \,\frac{\nabla{H}}{H}\times\bar{{\bf A}} \,.
\end{equation}
The last term in (\ref{ss}) is $H$-dependent. Therefore, we are left with a dynamical system (\ref{s}) determined by four functions, namely the three components of ${\bf A}$ and the Hamiltonian function $H$. 

Among the whole class (\ref{s}), an interesting subclass is provided by the choices
\begin{equation}
\label{ff}
\nabla\times{{\bf A}} = \lambda {\bf A} \,, \quad H = 1 \,,
\end{equation}
where $\lambda \in C^{\infty}(\mathbb{R}^{3},\mathbb{R})$ is an arbitrary smooth function . With the choice (\ref{ff}), we get 
\begin{equation}
\label{hff}
\dot{\bf r} = \lambda {\bf A} + {\bf A}\times\nabla\phi \,,
\end{equation}
so that the Hamiltonian vector field is immediately decomposed into a parallel and a perpendicular part to the vector field ${\bf A}$. 

As seen in (\ref{ff}), ${\bf A}$ is in the class of the force free vector fields \cite{Willis}. For instance, a force free field is given by the Arnold-Beltrami-Childress (ABC) vector field, 
\begin{equation}
\label{abc}
{\bf A} = (a\sin{z} + c\cos{y}, b\sin{x} + a\cos{z}, c\sin{y} + b\cos{x}) \,,
\end{equation}
for $a, b, c$ real non negative parameters. In this case, $\lambda = 1$. The ABC flow is known to be generically non integrable. For instance, for $a = b = c$ there is even an analytic proof of non-integrability \cite{Ziglin}. Inserting (\ref{abc}) into (\ref{hff}), we would obtain a perturbed ABC flow endowed with a Jacobi bracket structure. It would be interesting to investigate the integrability properties of such a system. 

To conclude, notice that the dynamical vector field $v_H$ associated to (\ref{s}) is not necessarily solenoidal, because
\begin{equation}
\nabla\cdot\,{\bf v}_H = \nabla\phi\cdot\nabla\times(H {\bf A}) \,,
\end{equation}
an expression which may be non vanishing. 

\subsection{Hamiltonian vector fields associated to rank $2$ Jacobi structures}
The class of Jacobi structures described in subsection III.B yields the following Hamilton equations,
\begin{equation}
\label{ep}
\dot{\bf r} = \nabla(\mu H)\times\nabla\psi - \mu H\nabla\xi_{1}\times\nabla\xi_2 \,.
\end{equation}
Without any loss of generality, for this class of dynamical systems we can set $H = 1$, thanks to the rescaling $\mu \rightarrow \mu H$. Hence, we adopt the choice $H = 1$ in the following. The dynamical system (\ref{ep}) contains four free functions, namely $\mu, \xi_{1}, \xi_{2}$ and $\psi(\xi_{1},\xi_{2})$. 

Notice that $\psi$ is a constant of motion if time-independent, because
\begin{equation}
\frac{d\psi}{dt} = -\mu\nabla\psi\cdot\nabla\xi_{1}\times\nabla\xi_2 = 0 \,.\
\end{equation}
Therefore, to obtain a Hamiltonian description of a dynamical system in $\mathbb{R}^3$ using a Jacobi structure of the extended Poisson type, one must know a first integral. This is a most restrictive condition, in comparison with the case of the previous subsection. 

Using (\ref{cas1}) and (\ref{cas2}) and taking $H = 1$, one can show that the Casimirs of the rank 2 Jacobi structures in $\mathbb{R}^3$ are preserved by the flow of the associated Hamiltonian vector field, 
\begin{equation}
{\bf v}_{H}(C) = 0 \,,
\end{equation}
where $C$ is any Casimir function. Since one can readily derive all Casimirs from the recipe described by equations (\ref{gam}) and (\ref{crc}), we conclude that a constant of motion $\psi$ plus a rank 2 Jacobi structure amounts to complete integrability, with the corresponding time-independent constants of motion being $\psi$ and $C$. 

To conclude, we observe that the Hamiltonian vector field ${\bf v}_H$ associated to (\ref{ep}) (with $H = 1$) is not necessarily solenoidal,
\begin{equation}
\nabla\cdot\,{\bf v}_H = -  \nabla\mu\cdot\nabla\xi_{1}\times\nabla\xi_2 \,,
\end{equation}
which may be non vanishing. 

\section{Conclusion}
The two admissible classes of Jacobi structures in $\mathbb{R}^3$ were constructed. For simplicity, neighborhood of points into domains $\Omega \subset \mathbb{R}^3$ where the vector field ${\bf A}$ vanish were not considered. In this sense the results of the work are not completely general. 

One of the reasons why Poisson structures are so ubiquitous rests in their applications for dynamical systems \cite{Kisisel}-\cite{Ay}. Certainly Jacobi structures have been not already applied at the same level in connection with dynamical systems. The results of Section IV shows the possible classes of Hamiltonian vector fields endowed with Jacobi structures in $\mathbb{R}^3$. This is a first step towards a more detailed examination of the possibilities opened by Jacobi structures for dynamical systems. For instance, one can ask about the existence of some kind of energy-Casimir method for Hamiltonian systems on Jacobi manifolds. In addition, notice that the inverse problem of the construction of a non trivial Jacobi bracket associated to a given dynamical system was not considered here. This seems to be a difficult task. For instance, due to the dependence of the $\phi$ function on ${\bf A}$, the equation (\ref{s}) is a nonlinear equation for ${\bf A}$ for a given dynamical vector field. However, the class of Hamiltonian equations in the subsection IV.A is possibly more interesting for applications, since one need not to obtain a first integral for it. In contrast, the Hamiltonian description at subsection IV.B have to be constructed in terms of a function $\psi$ which was shown to be a first integral for the dynamical system, if $\partial\psi/\partial\,t = 0$. Therefore, it is more difficult to derive Jacobi structures of the type of subsection IV.B, considering non integrable dynamical systems in $\mathbb{R}^3$.

Recently, there has been interest on Leibniz manifolds \cite{Ortega}, that is, manifolds endowed with brackets satisfying the derivation property but not necessarily the Jacobi identity. These Leibniz manifolds provide another possible alternative to generalize Poisson manifolds. It would be interesting to extend the results of the present work to Leibniz structures in $\mathbb{R}^3$. 
\vfill
\noindent{\bf Acknowledgments}\\
The author acknowledges the brazilian research funding agency Conselho Nacional de Desenvolvimento Cient\'{\i}fico e Tecnol\'ogico for financial support.

\end{document}